\documentclass[twocolumn,showpacs,preprintnumbers,amsmath,amssymb]{revtex4}

\usepackage{graphicx}
\usepackage{dcolumn}
\usepackage{bm}

\begin{document}

\preprint{APS/123-QCD}

\title{Dynamical chiral-symmetry breaking at $T=0$ and $T \ne 0$ \\
in the Schwinger-Dyson equation with lattice QCD data}

\author{H.~Iida, M.~Oka and H.~Suganuma}
\affiliation{\it Department of Physics, Tokyo Institute of Technology, 
Ohkayama 2-12-1, Meguro, Tokyo 152-8551, Japan}
\email{iida@th.phys.titech.ac.jp}

\date{\today}

\begin{abstract}
For the study of dynamical chiral-symmetry breaking (DCSB) in QCD, 
we investigate the Schwinger-Dyson (SD) formalism based on lattice QCD data.
From the quenched lattice data for the quark propagator in the Landau gauge, 
we extract the SD kernel function $K(p^2)$, 
which is the product of the quark-gluon vertex and the polarization factor in the gluon propagator, 
in an Ansatz-independent manner.
We find that the SD kernel function $K(p^2)$ exhibits infrared vanishing and 
a large enhancement at the intermediate-energy region around $p \sim 0.6{\rm GeV}$.
We investigate the relation between the SD kernel and the quark mass function, and find that 
the infrared and intermediate energy region $(0.35 {\rm GeV}<p<1.5{\rm GeV})$ would be relevant for DCSB. 
We apply the lattice-QCD-based SD equation to thermal QCD, 
and calculate the quark mass function at the finite temperature.
Spontaneously broken chiral symmetry is found to be restored at the high temperature above 100 MeV.
\end{abstract}

\maketitle

\section{\label{sec:Introduction}Introduction}

Quantum chromodynamics (QCD) has been accepted as the fundamental
theory of the strong interaction of hadrons~\cite{MGS}. Due to the
asymptotic freedom of QCD, the perturbative calculation is 
applicable to the high-energy process in the hadron reactions.
In contrast, low-energy QCD becomes a strong-coupling gauge theory, 
and exhibits interesting nonperturbative phenomena such as color confinement~\cite{conf2000} 
and dynamical chiral-symmetry breaking (DCSB)~\cite{NJL61,H8491,M8393}.

In perturbative QCD, the QCD scale parameter $\Lambda _{\rm QCD}$ is one of 
the most important quantities, and provides a typical scale of the strong interaction.
At the one-loop level, the running coupling constant is expressed as 
\begin{equation}
\alpha_s (p^2) \equiv \frac{g^2(p^2)}{4\pi}=
\frac{12\pi}{(11N_c-2N_f) \ln(p^2/\Lambda ^2_{\rm QCD})},
\label{eqn:runningcoupling}
\end{equation}
where $N_c$ and $N_f$ being the numbers of the color and flavor,
respectively~\cite{MGS}. 

The perturbative result is valid only at a large momentum scale $p^2$.
Indeed, the running coupling constant $\alpha_s(p^2)$ 
becomes large in the infrared region,
and formally diverges at $p^2 = \Lambda _{\rm QCD}^2$ in Eq.(\ref{eqn:runningcoupling}).
Accordingly, the simple perturbation is no more meaningful
and some nonperturbative effect should appear in this strong-coupling region.

As a quantitative aspect, the experimental data suggest 
$\alpha_s(M_Z^2) \simeq 0.1172\pm 0.002$ and 
the QCD scale parameter is estimated as 
$\Lambda_{\rm QCD}(\overline{\rm MS})=216 \pm 25{\rm MeV}$
in the $\overline{\rm MS}$ scheme with $N_f=5$~\cite{PDG02}.
From Eq.(\ref{eqn:runningcoupling}), one finds a rough estimation as  
$\alpha_s((0.5{\rm GeV})^2)\sim 1$, which means the breakdown 
of the validity of the perturbation theory at $p \sim 0.5{\rm GeV}$.
In other words, one may expect the appearance of nonperturbative 
effects due to the strong coupling 
in the infrared region as $p \sim 0.5 {\rm{GeV}}$ 

DCSB is one of the outstanding nonperturbative features in QCD.
The chiral symmetry, which the QCD lagrangian possesses in the massless quark limit, 
is spontaneously broken in the nonperturbative QCD vacuum~\cite{CL84}.
In the mathematical viewpoint, DCSB leads to the low-energy theorem and the current algebra~\cite{CL84,BKY88}, 
and,  in the phenomenological viewpoint, it results in the absence of the parity doubling for hadrons.
In terms of the nontrivial QCD vacuum, DCSB is characterized by the quark condensate 
$\langle \bar qq \rangle \simeq -(225 \pm 25 {\rm MeV})^3$~\cite{GL82},
which is caused by attractive interaction acting the quark-antiquark pair like 
Cooper-pair condensation in the superconductivity~\cite{NJL61}.
Such a physical process has been demonstrated by using the effective
models of QCD such as the Nambu-Jona-Lasinio model~\cite{NJL61,HK94}
and the instanton model~\cite{DP84}.
These models  suggest that the almost massless quark acquires a large effective mass of  
$M \simeq 300{\rm{MeV}}$ as a result of DCSB 
and behaves as a massive constituent quark~\cite{G82} in the infrared region.
The pion is identified as the Nambu-Goldstone boson 
associated with DCSB and obeys the low-energy theorem and the current algebra~\cite{CL84,BKY88}, 
where the pion decay constant, $f_\pi  \simeq 93 {\rm{MeV}}$, is
also a relevant quantity characterizing DCSB~\cite{CL84}.
In this way, DCSB is characterized by several quantities, 
the quark condensate $\langle \bar qq \rangle \simeq -(225 \pm 25{\rm{MeV}})^3$,
the effective quark mass $M \simeq 300{\rm{MeV}}$ and
the pion decay constant $f_\pi \simeq 93{\rm{MeV}}$.

As the study of DCSB based on QCD, the Schwinger-Dyson (SD) equation~\cite{H8491,M8393,HT02} 
and the Bethe-Salpeter equation~\cite{AKM9091}, which are expressed as nonlinear integral equations, 
have been used as popular methods to incorporate the infinite-order effects on the gauge coupling.
In most studies, however, the free gluon propagator was used in the SD equation, and 
possible nonperturbative effects in the infrared region were neglected. 
In several studies, the infrared nonperturbative effect was modeled and was taken into account in the SD equation~\cite{SST95,RW94}, 
but such modeling was done hypothetically.
Recently, the nonperturbative quark and gluon propagators have been 
calculated in lattice QCD Monte Carlo simulations with high accuracy.
In this paper, we try to combine the SD approach and the lattice QCD result for the quark and gluon propagators 
as the first step to understand the hadron physics in terms of quarks and gluons.

In section 2, we summarize the quark and the gluon propagators in lattice QCD. 
Using the lattice quark propagator, we calculate the pion decay constant $f_\pi$ in the Pagels-Stokar approximation, 
and the quark condensate $\langle\bar qq\rangle$. 
In section 3, we briefly review the SD formalism for the quark field in QCD.
In section 4, from the lattice QCD data of the quark propagator, 
we extract the SD kernel function, 
which is expressed by the product of the quark-gluon vertex and 
the polarization factor in the gluon propagator. 
In section 5, we investigate the relation between the SD kernel and the quark mass function through 
several tests with modified SD kernels.
In section 6, we apply the lattice-QCD based SD equation  
to the finite temperature system, and calculate the quark mass function 
and the quark condensate.
Section 7 is devoted to the summary and concluding remarks.

\section{\label{sec:action}Propagators of quarks and gluons in lattice QCD}

In this section, we summarize the recent lattice QCD results for the quark and the gluon propagators.
The lattice QCD Monte Carlo simulation is the first principle calculation of the strong 
interaction directly based on QCD in the Euclidean metric~\cite{R92MM94}.
In these years, relatively accurate lattice QCD calculations have been performed for the 
quark and gluon propagators in the Landau gauge at the quenched level~\cite{BBLWZ00,BHW02,BBLWZ02}.
These propagators are considered to include all the nonperturbative effects in quenched QCD.

In the Landau gauge, the Euclidean gluon propagator is generally expressed by  
\begin{eqnarray}
 D_{\mu\nu}(p^2)= \frac{d(p^2)}{p^2} \left( \delta_{\mu \nu} -\frac{p_{\mu} p_{\nu}}{p^2} \right),
\label{eqn:gluonprop}
\end{eqnarray}
where we refer to $d(p^2)$ as the ``polarization factor" in the gluon propagator.
 
We find that the lattice QCD data~\cite{BBLWZ00} for the polarization factor $d(p^2)$ 
is well described by the analytic function of 
\begin{eqnarray}
d(p^2)= Z \frac{p^4+ap^2}{p^4+b^2}
\label{eqn:polarization}
\end {eqnarray}
with $a \simeq 1.94 {\rm GeV}^2$ and $b \simeq 0.486 {\rm GeV}^2$, 
as shown in Fig.1.
The constant factor $Z \simeq 1.80$ denotes the wave-function renormalization of 
the gluon field at the lattice cutoff.

Note here that there are two remarkable features in the functional shape of $d(p^2)$.
\begin{enumerate}
\item
The polarization factor $d(p^2)$ exhibits the infrared vanishing and $d(p^2)$ is proportional to $p^2$ in the infrared region as $p < 0.5{\rm GeV}$.
\item
The polarization factor $d(p^2)$ exhibits a large enhancement in the intermediate-energy region as $p \sim 1{\rm GeV}$.
\end{enumerate}

Next, we summarize the quark propagator $S(p)$ in the Landau gauge in lattice QCD.
In general, the quark propagator in the Landau gauge is expressed as
\begin{eqnarray}
S(p)=\frac{Z(p^2)}{\not p+M(p^2)}
\label{eqn:quarkprop}
\end{eqnarray}
in the Euclidean metric.
Here, $M(p^2)$ is called as the quark mass function, and 
$Z(p^2)$ corresponds to the wave-function renormalization of the quark field~\cite{H8491,M8393}. 
In terms of the quark propagator, DCSB is characterized by 
the mass generation as $M(p^2) \neq 0$.

The quark mass function $M(p^2)$ in the Landau gauge 
is recently measured in lattice QCD at the quenched level~\cite{BHW02}.
Figure 2 shows the quark mass function $M(p^2)$ in the chiral limit 
obtained from lattice QCD at the quenched level~\cite{BBLWZ00}. 
The lattice data in the chiral limit can be fitted by  
\begin{eqnarray}
M(p^2)=\frac{M_0}{1+(p/\bar p)^\gamma}
\label{eqn:quarkmass}
\end{eqnarray}
with $M_0 \simeq$260 MeV, $\bar p \simeq$870MeV and $\gamma \simeq$3.04~\cite{BHW02},
although there is some ambiguity on the fit function. 
The infrared quark mass $M(0)=M_0\simeq 260{\rm MeV}$ seems consistent with the 
constituent quark mass in the quark model~\cite{G82}.

As for the quark wave-function renormalization factor $Z(p^2)$, the lattice QCD results seem somehow unclear~\cite{BHW02,BBLWZ02} 
due to the relatively large statistical and systematic errors. 
For large $p^2$, one finds $Z(p^2) \simeq 1$ in the Landau gauge, while lattice data seem to indicate $Z(p^2)<1$ for small $p^2$.
For the practical calculation in this paper, we adopt $Z(p^2) \simeq 1$ in the Landau gauge.

Using the lattice QCD result for the quark propagator, 
the quark condensate $\langle \bar qq \rangle$, which is the order parameter of DCSB, can be estimated 
from the trace of the quark propagator as 
\begin{eqnarray}
\langle \bar qq \rangle_\Lambda
&=&-\int^\Lambda  {d^4p \over (2\pi )^4}{\rm tr} S_q(p) \nonumber \\
&=&-{N_c \over 4\pi ^2}\int_0^{\Lambda ^2} dp^2 p^2 \frac{Z(p^2) M(p^2)}{p^2+M^2(p^2)}
\label{eqn:quarkcond}
\end{eqnarray}
at the renormalization point $\Lambda$.
With the assumption of $Z(p^2)=1$, 
one obtains $\langle \bar qq \rangle_{\rm \Lambda=1{\rm GeV}} \simeq -(223{\rm MeV})^3$, $\langle \bar qq \rangle_{\Lambda=4{\rm GeV}} \simeq -(300{\rm MeV})^3$ and 
$\langle \bar qq \rangle_{\Lambda=\infty} \simeq -(320{\rm MeV})^3$, 
which seem consistent with the standard value $\langle \bar qq \rangle \simeq -(225\pm 25 {\rm MeV})^3$. 

We calculate the pion decay constant $f_\pi$, which also characterizes DCSB, 
in the Pagels-Stokar approximation~\cite{PS7980},
\begin{eqnarray}
f_\pi ^2={N_c \over 4\pi ^2}\int_0^\infty &dp^2&
{p^2 M(p^2) \over \{p^2+M^2(p^2)\}^2} \nonumber \\
&\times& \left( M(p^2) -{p^2 \over 2}\cdot {dM(p^2) \over dp^2} \right),
\label{eqn:piondecayconst}
\end{eqnarray}
where $Z(p^2)=1$ is assumed.
With the lattice data for $M(p^2)$ in Eq.(\ref{eqn:quarkmass}), 
we obtain $f_\pi\simeq 87{\rm MeV}$, which is consistent with the experimental value $f_\pi \simeq 93 {\rm MeV}$ 
or the theoretical estimation of $f_\pi \simeq 87 {\rm MeV}$ in the chiral limit from the chiral perturbation~\cite{GL82}.

\section{The Schwinger-Dyson Formalism for Quarks}

The Schwinger-Dyson (SD) equation for the quark propagator $S(p)$ is described with 
the nonperturbative gluon propagator $D_{\mu\nu}(p)$ and 
the nonperturbative quark-gluon vertex $g \Gamma_\nu(p,q)$ as 
\begin{eqnarray}
S^{-1}(p)=S^{-1}_0 
+C_Fg^2 \int\frac{d^4 q}{(2\pi)^4} \gamma_\mu S(q) D_{\mu\nu}(p-q) \Gamma_\nu(p,q) \ 
\label{eqn:SDE1}
\end{eqnarray}
in the Euclidean metric.
Here, $S_0(p)$ denotes the bare quark propagator, and 
the color factor of quarks has been calculated as $C_F=4/3$.

In the practical calculation for QCD, however, the SD formalism is drastically truncated: 
the perturbative gluon propagator and the one-loop running coupling are used instead of 
the nonperturbative quantities in the original formalism.
This simplification seems rather dangerous because some of 
nonperturbative-QCD effects are neglected.

We formulate the SD equation for quarks in the chiral limit in the Landau gauge.
By taking the trace of Eq.(\ref{eqn:SDE1}), one finds 
\begin{eqnarray}
\frac{M(p^2)}{Z(p^2)}&=&\frac{C_Fg^2}{4} \int \frac{d^4q}{(2\pi)^4} 
D_{\mu\nu}(p-q) \nonumber \\
&\times& {\rm tr}\left\{\gamma_\mu \frac{Z(q^2)}{\not q+M(q^2)}\Gamma_\nu(p,q)\right\} 
\label{eqn:SDE2}
\end{eqnarray}
For the quark-gluon vertex, 
we assume the chiral-preserving vector-type vertex,  
\begin{eqnarray}
\Gamma_\mu(p,q)=\gamma_\mu\Gamma((p-q)^2),
\label{eqn:vertex}
\end{eqnarray}  
which keeps the chiral symmetry properly.
Note here that, to preserve the Ward-Takahashi identity for the axial vector vertex, 
the gluon momentum $(p-q)$ should be taken as the argument of the quark-gluon vertex $\Gamma$ 
in the ladder approximation of the SD and BS equations~\cite{KM92}. 
(In contrast, to be strict, the Higashijima-Miransky approximation~\cite{H8491,M8393} explicitly breaks 
the chiral symmetry in the formalism~\cite{KM92}.) 
Then, one obtains 
\begin{eqnarray}
\frac{M(p^2)}{Z(p^2)}
&=&C_Fg^2 \int\frac{d^4 q}{(2\pi)^4}
\frac{Z(q^2)M(q^2)}{q^2+M^2(q^2)} \nonumber \\
&\times& \Gamma((p-q)^2) D_{\mu\mu}((p-q)^2). 
\label{eqn:SDE3}
\end{eqnarray}
In the Landau gauge, the Euclidean gluon propagator takes the general form of 
\begin{eqnarray}
D_{\mu\nu}(p^2)= \frac{d(p^2)}{p^2} \left( \delta_{\mu \nu} -\frac{p_{\mu} p_{\nu}}{p^2} \right)
\label{eqn:gluonpropform}
\end{eqnarray}
with the gluon polarization factor $d(p^2)$.
Therefore, Eq.(\ref{eqn:SDE3}) is expressed as 
\begin{eqnarray}
\frac{M(p^2)}{Z(p^2)}&=&3C_Fg^2 \int\frac{d^4 q}{(2\pi)^4} \frac{Z(q^2)M(q^2)}{q^2+M^2(q^2)}
\nonumber \\
&\times& \frac{\Gamma((p-q)^2)d((p-q)^2)}{(p-q)^2}. 
\label{eqn:SDE4}
\end{eqnarray}
Here, we define the kernel function  
\begin{eqnarray}
K(p^2) \equiv g^2 \Gamma(p^2) d(p^2)
\label{eqn:kernel}
\end{eqnarray}
as the product of the quark-gluon vertex $\Gamma(p^2)$ and the gluon polarization factor $d(p^2)$.
Then, the SD equation is rewritten as   
\begin{eqnarray}
\frac{M(p^2)}{Z(p^2)}=3C_F \int\frac{d^4 q}{(2\pi)^4}  \frac{Z(q^2)M(q^2)}{q^2+M^2(q^2)} \frac{K((p-q)^2)} {(p-q)^2}.
\label{eqn:SDE5}
\end{eqnarray}
Note that the actual kernel $\hat K_{\rm SD}(p^2)$ of the SD equation 
includes the Coulomb-propagator factor $1/p^2$ as
\begin{eqnarray}
\hat K_{\rm SD}(p^2) \equiv \frac{K(p^2)}{p^2}= \frac{g^2 \Gamma(p^2)d(p^2)}{p^2}.
\label{eqn:SDkernel}
\end{eqnarray}

In the Landau gauge, from several theoretical arguments~\cite{H8491,AKM9091,KM92},  
we expect that the quark wave-function renormalization is not so significant, 
and therefore we approximate as $Z(p^2)=1$.
In this approximation, the SD equation reduces to 
\begin{eqnarray}
M(p^2)=3C_F \int\frac{d^4 q}{(2\pi)^4}  \frac{M(q^2)}{q^2+M^2(q^2)} \frac{K((p-q)^2)} {(p-q)^2}.
\label{eqn:SDE6}
\end{eqnarray}
In the following sections, we analyze this type of the SD equation.

\section{Extraction of the Kernel Function in the SD Equation from Lattice QCD}

In this section, we extract the kernel function $K(p^2)\equiv g^2 \Gamma(p^2)d(p^2)$ in the SD equation (\ref{eqn:SDE6}) 
from the quark mass function $M(p^2)$ obtained in lattice QCD in an Ansatz-independent manner. 
By shifting the integral variable from $q$ to $\tilde q \equiv p-q$, 
we rewrite 
Eq.(\ref{eqn:SDE6}) as 
\begin{eqnarray}
M(p^2)=3C_F \int\frac{d^4 \tilde q}{(2\pi)^4} \frac{M((p-\tilde q)^2)}{(p-\tilde q)^2+M^2((p-\tilde q)^2)} \frac{K(\tilde q^2)} {\tilde q^2}. 
\label{eqn:SDE7}
\end{eqnarray}
Therefore, we obtain 
\begin{eqnarray}
M(p^2)=\int_0^\infty d \tilde q^2 \Theta(p^2, \tilde q^2) K(\tilde q^2), 
\label{eqn:SDE8}
\end{eqnarray}
where $\Theta(p^2,q^2)$ is defined with  $M(p^2)$ as 
\begin{widetext}
\begin{eqnarray}
\Theta(p^2,q^2) \equiv \frac{3C_F}{8\pi^3}  \int_0^\pi d\theta \sin^2 \theta 
\frac{M(p^2+q^2-2pq\cos\theta)}{p^2+q^2-2pq\cos\theta+M^2(p^2+q^2-2pq\cos\theta)}. 
\label{eqn:Theta}
\end{eqnarray}
\end{widetext}
Regarding the momentum squared ($p^2$, $\tilde q^2$) as suffices ($m$, $n$), Eq.(\ref{eqn:SDE8}) can be rewritten as 
\begin{eqnarray}
M_m= \sum_{n} \Theta_{mn} K_n.
\label{eqn:SDE9}
\end{eqnarray}
Here, $\Theta_{mn}$ is a real symmetric matrix on $m$ and $n$ as
\begin{eqnarray}
\Theta_{mn}=\Theta_{nm} \in {\bf R},
\end{eqnarray}
because of $\Theta(p^2,q^2)=\Theta(q^2,p^2) \in {\bf R}$.

Once the quark mass function $M(p^2)$ is obtained, $\Theta(p^2,q^2)$ is calculable with Eq.(\ref{eqn:Theta}), and, 
using Eq.(\ref{eqn:SDE9}), we can extract $K_n$ directly from $\Theta_{mn}$ and $M_n$ as  
\begin{eqnarray}
K_m= \sum_{n} \Theta^{-1}_{mn} M_n.
\label{eqn:SDE10}
\end{eqnarray}
Since the quark mass function $M(p^2)$ is given by Eq.(\ref{eqn:quarkmass}) 
in lattice QCD, we can calculate the kernel function $K(p^2)$ from Eq.(\ref{eqn:SDE10}) 
without any assumption of the functional form on the kernel function $K(p^2)$.
For the practical calculation, 
we discretize the momentum squared $p^2$ and $\tilde q^2$ 
after the proper transformation as $p^2=\arctan \alpha$ and $\tilde q^2=\arctan \beta$, 
and solve Eq.(\ref{eqn:SDE9}) for $K_n$.

As shown in Fig.3, we numerically extract the kernel function $K(p^2)\equiv g^2 \Gamma(p^2)d(p^2)$ 
from the lattice QCD result, Eq.(\ref{eqn:quarkmass}), for the quark propagator in the Landau gauge.
For the check of the validity, we have confirmed that the mass function $M(p^2)$ in 
Eq.(\ref{eqn:quarkmass}) is precisely reproduced with the obtained kernel function $K(p^2)$.

As remarkable features, 
we find ``infrared vanishing" and ``intermediate enhancement" 
in the kernel function $K(p^2)$ in the SD equation~\cite{IOS03}:
\begin{enumerate}
\item
The SD kernel function $K(p^2)$ seems consistent with zero in the very infrared region as 
\begin{eqnarray}
K(p^2 < 0.1{\rm GeV}^2) \simeq 0.
\label{eqn:IRvanishing}
\end{eqnarray}
\item
The SD kernel function $K(p^2)$ exhibits a large enhancement in the intermediate-energy region around $p \sim$ 0.6GeV.
In fact, $K(p^2)$ takes the maximal value $K_{\rm max}=38.76$ at $p^2\simeq0.368{\rm GeV}^2 \simeq(0.607 {\rm GeV})^2$.
\end{enumerate}

These tendencies of infrared vanishing and intermediate enhancement in the kernel function 
$K(p^2)\equiv g^2 \Gamma(p^2)d(p^2)$ 
are qualitatively observed also in the direct lattice-QCD measurement for   
the polarization factor $d(p^2)$ in the gluon propagator 
in the Landau gauge~\cite{BBLWZ00} as shown in Fig.1.

Note here that the usage of the perturbative gluon propagator and the one-loop running coupling $g(p^2)$
corresponds to the perturbative SD kernel function
\begin{eqnarray}
K_{\rm pert}(p^2)=g^2(p^2)=\frac{48\pi^2}{(11N_c-2N_f) \ln(p^2/\Lambda^2_{\rm QCD})}.
\label{eqn:pertkernel}
\end{eqnarray}
For comparison, we add by the dotted curve in Fig.3 the perturbative SD kernel function $K_{\rm pert}(p^2)$ at the one-loop level.
Both in the infrared and in the intermediate energy regions,
$K_{\rm pert}(p^2)$ largely differs from the present result $K(p^2)$ based on lattice QCD.
In fact, the simple version of the SD equation using the perturbative gluon propagator and the one-loop running coupling 
would be too crude for the quantitative study of QCD.

\section{Relation between DCSB and the SD kernel}

In this section, we investigate the relation between the quark mass function $M(p^2)$ 
and the SD kernel function $K(p^2)$ through several simple tests, 
considering the relevant momentum region for DCSB.
Since the SD equation is a nonlinear equation, it is difficult to find out such relevant region for DCSB in a rigorous manner.
In order to determine the relevant momentum region for DCSB, we calculate the quark mass functions for modified SD kernels and 
compare the results with the original one.
We consider the following four modifications.
\begin{enumerate}
\item An ultraviolet (UV) cutoff for the SD kernel
\item An infrared (IR) cutoff for the SD kernel 
\item An intermediate (IM) suppression for the SD kernel
\item A simple scaling of the SD kernel
\end{enumerate}

\subsection{Ultraviolet cut case}

To check the relevant momentum region for DCSB, 
we investigate the ultraviolet (UV) cut case, infrared (IR) cut case and 
intermediate (IM) suppression case for the SD kernel,
To begin with, we investigate the UV-cut case for the SD equation (\ref{eqn:SDE6}) 
by using the UV-cut SD kernel function, 
\begin{eqnarray}
K_{\rm UV}(p^2; \Lambda_{\rm UV}) \equiv K(p^2)\theta(\Lambda_{\rm UV}^2-p^2),
\label{eqn:UVcut}
\end{eqnarray}
instead of $K(p^2)$.

For each $\Lambda_{\rm UV}$, we solve the UV-cut SD equation and obtain the corresponding solution 
$M_{\rm UV}(p^2;\Lambda_{\rm UV})$ for the quark mass function.
We show in Fig.4(a) the infrared quark mass $M_{\rm UV}(0;\Lambda_{\rm UV})$ plotted against $\Lambda_{\rm UV}$.
For $\Lambda_{\rm UV} > 2{\rm GeV}$, 
almost no effect is observed as $M_{\rm UV}(p^2; \Lambda_{\rm UV}) \simeq M_{\rm UV}(p^2; \infty)=M(p^2)$, which clearly 
indicates that UV region is not important for DCSB.
In contrast, a significant reduction of the quark mass function $M_{\rm UV}(p^2; \Lambda_{\rm UV})$ 
is observed for $\Lambda_{\rm UV}<1.5{\rm GeV}$.
In particular, for $\Lambda_{\rm UV} < \Lambda_{\rm UV}^{\rm crit}  \simeq 0.9{\rm GeV}$, 
no DCSB is observed as $M_{\rm UV}(p^2; \Lambda_{\rm UV})=0$, which may suggest that 
the momentum scale of $p \sim 1 {\rm GeV}$ plays an important role for DCSB.
Figure 4(b) shows the UV cut SD kernel function for $\Lambda_{\rm UV}^{\rm crit}$, which is the critical value on DCSB.

This result seems natural because the strong coupling nature at the infrared and intermediate energy regions 
would be essential for DCSB in QCD \cite{SST95,BPRT03}.

\subsection{Infrared cut case}

To investigate the role of the infrared (IR) region for DCSB, we consider also the IR cut case 
by using the IR-cut SD kernel function, 
\begin{eqnarray}
K_{\rm IR}(p^2;\Lambda_{\rm IR})\equiv K(p^2)\theta(p^2-\Lambda_{\rm IR}^2),
\label{eqn:IRcut}
\end{eqnarray}
instead of $K(p^2)$ in Eq.(\ref{eqn:SDE6}).
For each $\Lambda_{\rm IR}$, we solve the IR-cut SD equation and obtain 
the corresponding solution $M_{\rm IR}(p^2;\Lambda_{\rm IR})$.
We show in Fig.5(a) the infrared quark mass $M_{\rm IR}(0;\Lambda_{\rm IR})$ plotted against the IR cutoff $\Lambda_{\rm IR}$.

For $\Lambda_{\rm IR}<$0.35GeV, no significant effect is observed for DCSB, according to the infrared vanishing of the SD kernel.
In contrast, for $\Lambda_{\rm IR} > \Lambda_{\rm IR}^{\rm crit} \simeq 0.53{\rm GeV}$, 
no DCSB is observed as $M_{\rm IR}(p^2;\Lambda_{\rm IR})=0$.
These results seem to indicate the relevant role of the infrared region as $0.35{\rm GeV}<p<0.53{\rm GeV}$ for DCSB.
Figure 5(b) shows the IR-cut SD kernel function for $\Lambda_{\rm IR}^{\rm crit}$, which is the critical value on DCSB.

\subsection{Intermediate suppressed case}

The SD kernel function $K(p^2)$ obtained from the lattice QCD data of the quark propagator  
indicates the intermediate enhancement, 
and takes a maximal value $K_{\rm max}=38.76$ at $p^2 \simeq 0.368 {\rm GeV}^2 \simeq (0.607 {\rm GeV})^2$ 
in the present analysis. (See Fig.3)
To investigate the role of the intermediate (IM) enhancement of the SD kernel for DCSB,  
we examine the IM-suppressed SD kernel function, 
\begin{eqnarray}
K_{\rm IM}(p^2;c)\equiv {\rm Min} (K(p^2), cK_{\rm max}),
\label{IMsuppress}
\end{eqnarray}
with a real and positive constant $c$.

For each $c$, we calculate the quark mass function $M_{\rm IM}(p^2; c)$ 
using the SD equation (\ref{eqn:SDE6}) with the modified kernel $K_{\rm IM}(p^2;c)$.
We show in Fig.6(a) the infrared quark mass $M_{\rm IM}(0; c)$ plotted against $c$.
Of course, for $c \ge 1$, one finds $K_{\rm IM}(p^2;c)=K(p^2)$ and $M_{\rm IM}(p^2;c)=M(p^2)$.
As $c$ decreases from 1, $M_{\rm IM}(p^2;c)$ rapidly decreases, 
and no DCSB is found as $M_{\rm IM}(p^2;c)=0$ for $c<c_{\rm crit} \simeq 0.58$.
Figure 6(b) shows the IM-suppressed SD kernel function for $c_{\rm crit}$, which is the critical value on DCSB.
This would indicate that the intermediate enhancement of the SD kernel in the region of 
$0.2{\rm GeV}^2<p^2<0.8{\rm GeV}^2$,  i.e., $0.4{\rm GeV}<p<0.9{\rm GeV}$, plays an important role for DCSB.

From the above analyses in Subsect. A, B and C, 
the relevant momentum region for DCSB is considered to be
the IR and the IM regions as 
\begin{eqnarray}
0.35{\rm GeV} < p < 1.5{\rm GeV}.
\label{eqn:relevantregion}
\end{eqnarray}

\subsection{Sensitivity of DCSB on the magnitude of the SD kernel}

In general, DCSB is realized when the SD kernel is large enough.
In fact, the strong-coupling nature is expected to be the origin of DCSB, 
as was first pointed out in the Nambu-Jona-Lasino (NJL) model~\cite{NJL61}, 
where DCSB is realized for the strong coupling case, e.g., $g_{\rm NJL} \Lambda^2>\pi N_c$ for 
the quark version of the NJL model~\cite{H8491,M8393,HK94}. 

Finally in this section, we consider a simple scaling of the SD kernel 
with a real and positive constant $\lambda$. 
In fact, we solve the modified SD equation with 
the scaled SD kernel function 
\begin{eqnarray}
K(p^2; \lambda) \equiv \lambda K(p^2),
\label{eqn:scaled}
\end{eqnarray}
and investigate the corresponding quark mass function $M(p^2;\lambda)$.
This modification can be physically interpreted as the check on the sensitivity of DCSB 
on the gauge coupling magnitude, since the $K(p^2)$ is proportional to $g^2$.
In fact, the SD-kernel scaling as Eq.(\ref{eqn:scaled}) is equivalent to 
the scaling of the QCD gauge coupling $g^2$ as 
\begin{eqnarray}
g_\lambda^2 \equiv \lambda g^2
\label{eqn:scaledcoupling}
\end{eqnarray}
in the SD equation.
(From Eq.(\ref{eqn:SDE5}), this simple SD-kernel scaling can be also interpreted as the introduction of 
the constant quark wave-function renormalization $Z$ as $\lambda =Z^2$, 
although we have ignored its effect as $Z=1$.)
 
For each $\lambda$, we solve the SD equation (\ref{eqn:SDE6}) with the scaled kernel function $K(p^2; \lambda)$, 
and obtain the corresponding quark mass function $M(p^2;\lambda)$.
We show in Fig.7 the infrared quark mass $M(0;\lambda)$ against the scale parameter $\lambda$.
For $\lambda \ge 0.8$, the quark mass function $M(p^2;\lambda)$ is an increasing function on $\lambda$ for fixed value of $p^2$.
For $\lambda < \lambda_{\rm crit} \simeq 0.8$, no DCSB occurs as $M(p^2; \lambda) = 0$.

This result indicates that the physical case of $\lambda=1$ may locate 
near the critical case $\lambda_{\rm crit} \simeq 0.8$ for DCSB. 
In accordance with the closeness to the critical case, 
the solution for the quark mass function seems rather sensitive on the overall scaling of the SD kernel near $\lambda=1$. 
This may be rather serious for the SD approach because, for more realistic calculation, we need more sophisticated treatment 
on the wave function renormalization, the vertex and so on.
Indeed, a similar approach in Ref.\cite{BPRT03} results in a smaller value for the pion decay constant $f_\pi$ from  
the BS equation with $Z(p^2)$ taken into account.

In this way, DCSB such as the quark mass function may be sensitive to 
the overall scaling of the SD kernel.
Our result indicates that DCSB is near the critical point as $\lambda_{\rm crit} \simeq 0.8$, i.e., 
$g^2 \simeq 1.25g_{\rm crit}^2$ for the coupling constant from Eq.(\ref{eqn:scaledcoupling}).

\ 

\section{Chiral Symmetry at Finite Temperature}

Finally, we demonstrate a simple application of 
the lattice-QCD-based SD equation to chiral symmetry restoration in finite-temperature QCD, using the Matsubara imaginary-time formalism. 

In this formalism, the quark field $q({\bf x}, \tau)$ and the gluon field $A_\mu({\bf x}, \tau)$ at a finite temperature $T$
obey the anti-periodic and periodic boundary condition in the imaginary-time direction, respectively, as 
\begin{eqnarray}
q({\bf x},\tau+1/T)&=&-q({\bf x},\tau),  \nonumber \\
A_\mu({\bf x},\tau+1/T)&=&A_\mu({\bf x},\tau), 
\label{eqn:periodic}
\end{eqnarray}
Accordingly, the temporal momentum variable $p_0$ is discretized to be the Matsubara frequency,
\begin{eqnarray}
p_0&\rightarrow&(2n+1)\pi T \equiv \omega_n \quad \hbox{for quarks},  \nonumber \\
p_0&\rightarrow&2n\pi T  \qquad \qquad \qquad \hbox{for gluons},
\label{eqn:Matsubarafreq}
\end{eqnarray}
and the corresponding integration over $p_0$ becomes the summation over the Matsubara frequencies as
\begin{eqnarray}
\int_{-\infty}^{\infty}\frac{dp_0}{2\pi}&\rightarrow& T \sum_{n=-\infty}^\infty 
\label{eqn:summation}
\end{eqnarray}
in the SD equation.

As a result, the SD equation for the thermal quark mass function $M_T({\bf p}^2,\omega_n^2)$ 
of the Matsubara frequency $\omega_n \equiv (2n+1)\pi T$ is given as 
\begin{widetext}
\begin{eqnarray}
M_T({\bf p}^2,\omega_n^2)=3C_F T \sum_{m=-\infty}^\infty \int \frac{d^3q}{(2\pi)^3} 
\frac{M_T({\bf q}^2,\omega_m^2)}{\omega_m^2+{\bf q}^2+M_T^2({\bf q}^2,\omega_m^2)} 
\frac{K((\omega_n-\omega_m)^2+({\bf p}-{\bf q})^2)}
{(\omega_n-\omega_m)^2+({\bf p}-{\bf q})^2}.
\label{eqn:SDET}
\end{eqnarray}
\end{widetext}
Here, $M_T({\bf p}^2,\omega_n^2)$ depends only on ${\bf p}^2$ and $\omega_n^2$, 
because of the three-dimensional rotational invariance and 
the imaginary-time reversal invariance of the system.

Using the kernel function $K(p^2)$ obtained in section IV, 
we solve the thermal SD equation (\ref{eqn:SDET}) for $M_T({\bf p}^2,\omega_n^2)$.
Figure 8 shows the numerical result for 
the thermal infrared quark mass $M_T({\bf p}^2=0,\omega_0^2)$ 
plotted against the temperature $T$.
Chiral symmetry restoration is found at 
a critical temperature $T_c \simeq 100{\rm MeV}$, which seems rather small in comparison 
with the critical temperature $T_c= 260 - 280{\rm MeV}$ of the quenched QCD phase transition~\cite{ISM02}.

In this calculation, the chiral phase transition is found to be of the second order, and 
we find near $T_c$ 
\begin{eqnarray}
M_T(0,\omega_0^2) \sim (1-T/T_c)^{0.47}
\label{eqn:criticalbehavior}
\end{eqnarray}
as the critical behavior,
although more careful treatments would be necessary for the argument of the critical phenomena.

Figure 9 shows the thermal quark mass function $M_T({\bf p}^2,\omega_n^2)$ at 
various Matsubara frequencies at a low temperature $T = 0.6T_c = 60 {\rm MeV}$ and 
at a high temperature $T= 0.9T_c =90{\rm MeV}$.
One finds the following tendencies for the thermal quark mass $M_T({\bf p}^2,\omega_n^2)$.
\begin{enumerate}
\item
At fixed $T$, 
$M_T({\bf p}^2,\omega_n^2)$ is a decreasing function of ${\bf p}^2$ for each Matsubara frequency $n$, 
and $M_T({\bf p}^2,\omega_n^2)$ decreases with $\omega_n^2$ for each value of ${\bf p}^2$.
\item
For fixed $n$ and ${\bf p}^2$, $M_T({\bf p}^2,\omega_n^2)$ decreases with the temperature $T$.
\end{enumerate}

As an interesting dependence of $\omega_n^2$ and ${\bf p}^2$, 
we find at each $T$ the ``covariant-like relation" \cite {SST96} for the thermal quark mass function as 
\begin{eqnarray}
M_T({\bf p}^2,\omega_n^2) \simeq \tilde M_T({\bf p}^2+\omega_n^2)= \tilde M_T(\hat p^2).
\label{eqn:covariantlike}
\end{eqnarray}
Here, $\hat p^2 \equiv {\bf p}^2+\omega_n^2={\bf p}^2+\{(2n+1)\pi T\}^2$ corresponds to 
the four-dimensional momentum squared, $p^2={\bf p}^2+p_0^2$, and actually reduces into $p^2$ at $T=0$.
For the demonstration of this relation, 
we define $\tilde M_T(\hat p^2)=M_T({\bf p}^2,\omega_0^2)$ and compare $\tilde M_T(\hat p^2)$ with 
$M_T({\bf p}^2,\omega_n^2)$ for each $n$ in Fig.10. We find an approximate coincidence between them  
at each $n$ even at temperature close to $T_c$.

The quark condensate at finite temperature is calculated with $M_T({\bf p}^2,\omega_n^2)$ as~\cite{SST96}
\begin{eqnarray}
&& \langle \bar qq(T) \rangle_\Lambda
=-T\sum_{n} \int^\Lambda {d^3p \over (2\pi )^3}{\rm tr} S_q(p) \\
&\equiv&-{2N_c \over \pi^2} T \sum_{n=-N}^N \int_0^{\sqrt{\Lambda^2-\omega_n^2}} dp p^2 
{M_T({\bf p}^2,\omega_n^2) \over \omega_n^2+{\bf p}^2+M_T^2({\bf p}^2,\omega_n^2)} \nonumber \\
&=&-{2N_c \over \pi^2} T \sum_{n=-N}^N \int_{\omega_n}^{\Lambda} d\hat p \hat p \sqrt{\hat p^2-\omega_n^2} 
{M_T(\hat p^2-\omega_n^2,\omega_n^2) \over \hat p^2+M_T^2(\hat p^2-\omega_n^2,\omega_n^2)}, \nonumber 
\label{eqn:quarkcondT}
\end{eqnarray}
where $N \equiv [\frac{\Lambda}{2\pi T}-\frac12] $ is the UV cutoff on the Matsubara frequency. 
Note that $N$ satisfies $(2N+1)\pi T \simeq \Lambda$, and corresponds to the momentum cutoff $\Lambda$.
In the limit of $T \rightarrow 0$, $\langle \bar qq(T) \rangle_\Lambda$ 
becomes the quark condensate $\langle \bar qq \rangle_\Lambda$ in Eq.(\ref{eqn:quarkcond}) at the zero temperature~\cite{SST96}.
We show the thermal quark condensate $\langle \bar qq(T) \rangle_\Lambda$
plotted against the temperature $T$ in Fig.11, which exhibits  
the critical behavior near $T_c$ as 
\begin{eqnarray}
\langle \bar qq(T) \rangle_{\Lambda=4{\rm GeV}} \sim (1-T/T_c)^{0.74}.
\label{eqn:criticalcondensate}
\end{eqnarray}

In this calculation, 
we have included only the (anti-)periodicity in the imaginary-time direction, and 
have ignored the nontrivial thermal effects on the quark and gluon propagators and vertex functions.
Nevertheless, we observe chiral symmetry restoration at high temperature 
and obtain a rough estimate of the critical temperature.
For more consistent calculation, it would be interesting to use the lattice QCD results on  
the quark and gluon propagators at finite temperature. 

\section{Summary and Concluding Remarks}

We have investigated the Schwinger-Dyson (SD) formalism based on lattice QCD data, and 
have studied dynamical chiral-symmetry breaking (DCSB) in QCD, 
We have extracted the SD kernel function $K(p^2)$, 
which is the product of the quark-gluon vertex and the polarization factor in the gluon propagator, 
in an Ansatz-independent manner from the quenched lattice data for the quark  propagator in the Landau gauge. 
We have found that the SD kernel $K(p^2)$ exhibits infrared vanishing and a large enhancement at the intermediate-energy region 
around $p \sim 0.6{\rm GeV}$.

We have investigated the relation between the quark mass function and the SD kernel, considering the important scale region for DCSB.
Several checks have indicated that 
the infrared and the intermediate energy regions as $0.35 {\rm GeV}<p<1.5{\rm GeV}$
would be relevant for DCSB. 
The ``intermediate kernel enhancement" is found to be also important for DCSB. 

We have applied the lattice-QCD-based SD equation to thermal QCD, 
and have calculated the quark mass function at the finite temperature.
We have found that spontaneously broken chiral symmetry is  
restored at the high temperature above 100 MeV.

In this paper, we have investigated the SD equation in the Landau gauge at the quenched level.
It is interesting to study the SD equation in other choices of the gauge such as the maximally Abelian gauge~\cite{AS99}, 
which may reveal possible relations between DCSB and color confinement~\cite{SST95}. 
It is also interesting to investigate the SD equation based on the quark and the gluon propagators in full QCD.  
As a next possible step, the application to finite density QCD would be an interesting subject, 
which is rather hard to be performed within the present lattice-QCD ability.

In near future, we expect that 
the combination of lattice QCD and the nonperturbative formalism such as the SD and the BS equations 
provides us a powerful tool to clarify nonperturbative aspects of the hadron physics based on QCD. 
In this framework, if one obtains accurate lattice QCD data, one would derive more reliable results. 
Therefore, in order to establish this new framework, it is much desired to obtain 
more accurate data on the quark and gluon propagators as well as vertex functions in lattice QCD.

\begin{acknowledgments}
We would like to thank Dr.~Y.~Sumino for his useful comments and discussions.
This work was supported in part by Grant for Scientific Research 
(No.11640261, No.12640274) from the Ministry of Education, 
Culture, Sports, Science and Technology, Japan. 
H.I. was supported by a 21st Century COE Program at
Tokyo Institute of Technology ``Nanometer-Scale Quantum Physics" by the
Ministry of Education, Culture, Sports, Science and Technology.
\end{acknowledgments}

\begin{figure}
\begin{center}
 \rotatebox{90}{
	\includegraphics[width=6cm,clip]{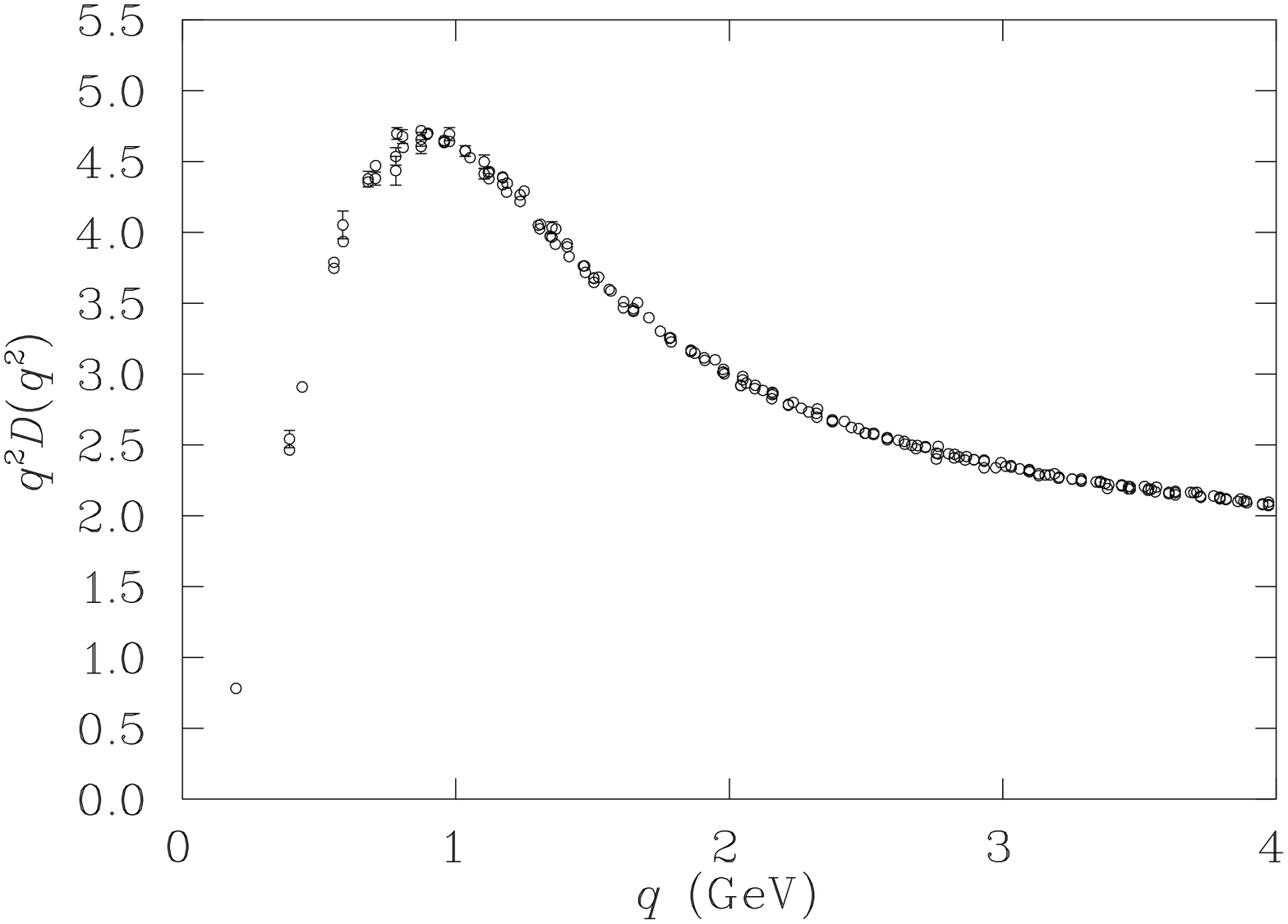}}
\end{center}
\caption{
The lattice QCD result taken from Ref.\cite{BBLWZ00} 
for the polarization factor $d(q^2) = q^2D(q^2)$ in the gluon propagator 
in the Landau gauge.
}
\label{GluonPropagator}
\end{figure}

\begin{figure}
\begin{center}
 \rotatebox{90}{
	\includegraphics[width=6cm,clip]{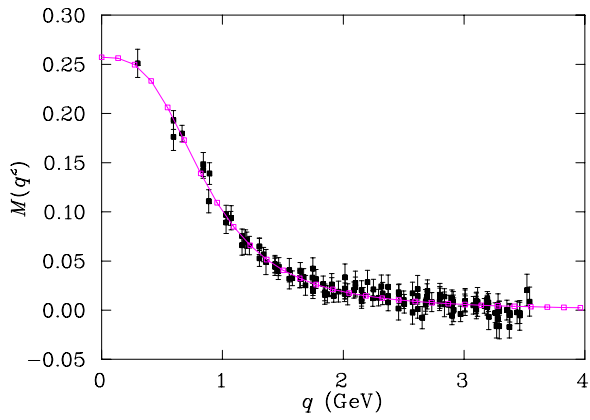}}
\end{center}
\caption{
The lattice QCD result taken from Ref.\cite{BHW02} 
for the quark mass function $M(q^2)$ 
in the chiral limit in the Landau gauge.
}
\label{QuarkPropagator}
\end{figure}

\begin{figure}
\begin{center}
 \rotatebox{-90}{
	\includegraphics[width=6cm,clip]{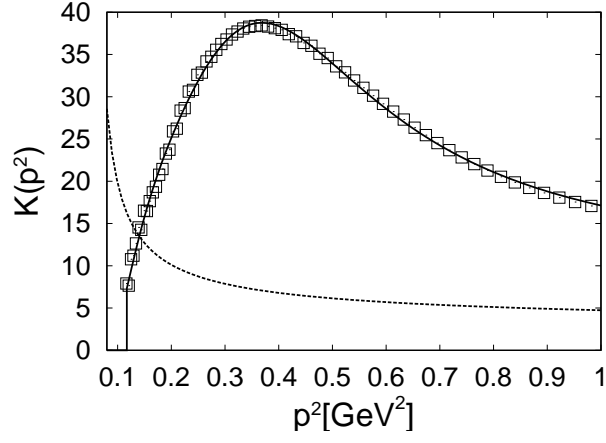}}
\end{center}
\caption{
The kernel function in the SD equation, $K(p^2) \equiv g^2 \Gamma(p^2)d(p^2)$, 
extracted from the lattice QCD result of the quark propagator in the Landau gauge. 
The calculated data are denoted by the square symbols, and  
the solid curve denotes a fit function for them.
The dotted curve denotes the perturbative SD kernel function $K_{\rm pert}(p^2)$ for comparison.
As remarkable features, $K(p^2)$ exhibits infrared vanishing and intermediate enhancement.
}
\label{SDKernelFunction}
\end{figure}

\begin{figure}
\begin{center}
 \rotatebox{-90}{
	\includegraphics[width=6cm,clip]{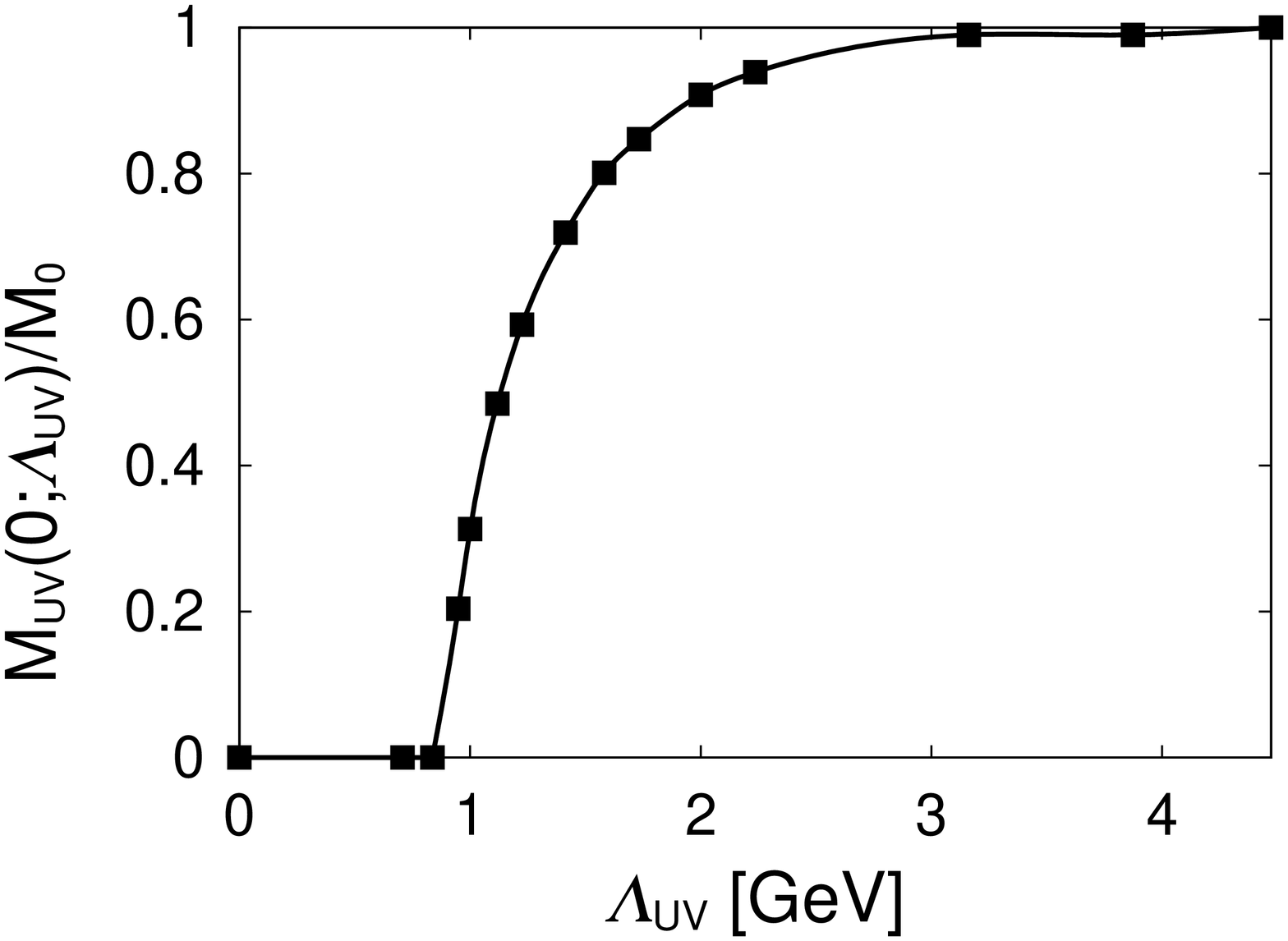}
	\includegraphics[width=6cm,clip]{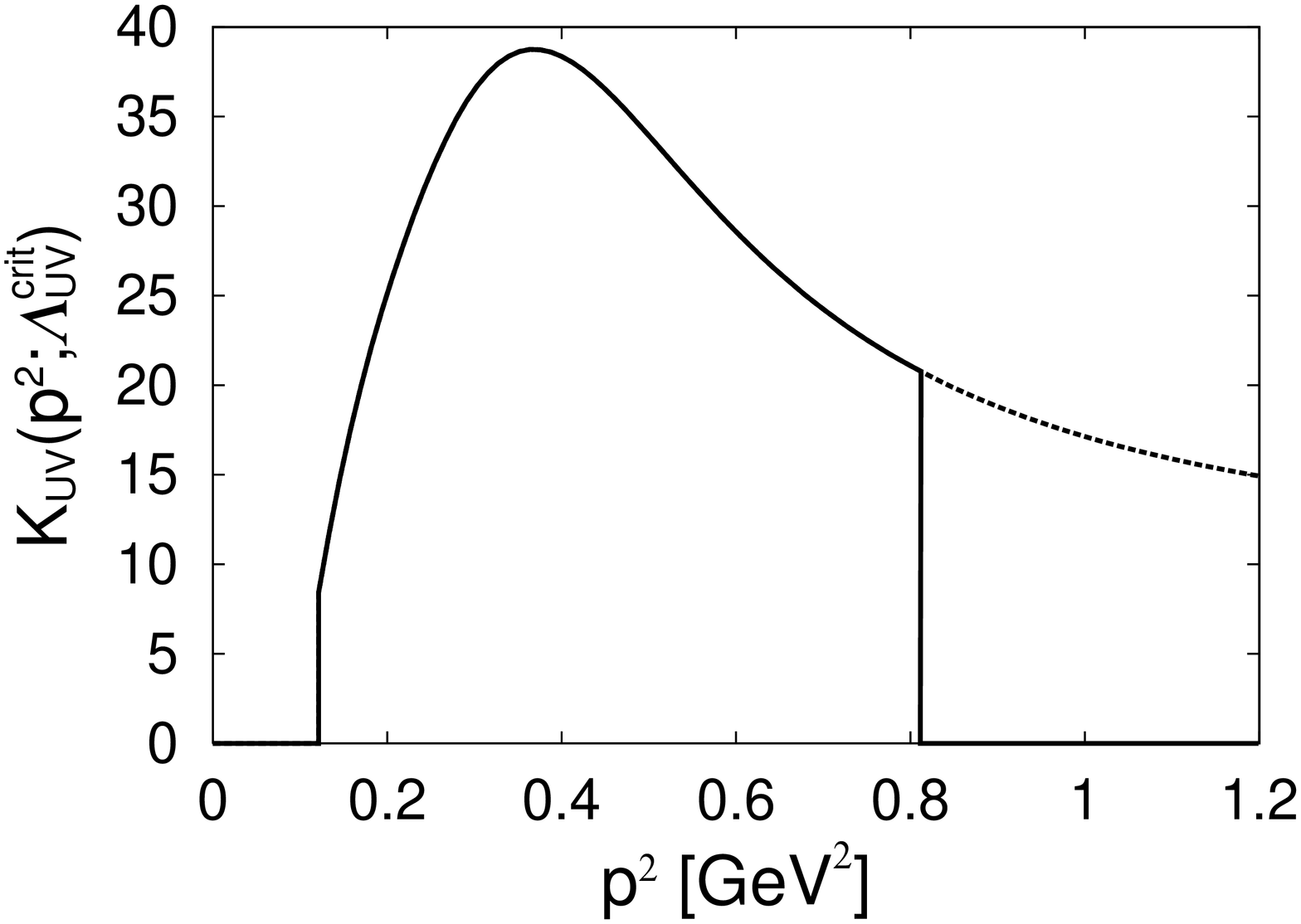}}
\end{center}
\caption{
(a) The infrared quark mass $M_{\rm UV}(0;\Lambda_{\rm UV})/M_0$ in the UV-cut SD equation 
plotted against the artificial UV-cutoff parameter $\Lambda_{\rm UV}$. $M_0 \simeq 260{\rm MeV}$ denotes the infrared quark mass. 
For $\Lambda_{\rm UV} > 2{\rm GeV}$, almost no effect is observed.
For $\Lambda_{\rm UV} < \Lambda_{\rm UV}^{\rm crit}  \simeq 0.9{\rm GeV}$, 
no DCSB is observed as $M_{\rm UV}(p^2; \Lambda_{\rm UV})=0$.
(b) The UV-cut SD kernel $K_{\rm UV}(p^2;\Lambda_{\rm UV}^{\rm crit})$ for the critical case on DCSB.
The dotted curve denotes the original SD kernel function $K(p^2)$.
}
\label{UVcutcase}
\end{figure}

\begin{figure}
\begin{center}
 \rotatebox{-90}{
	\includegraphics[width=6cm,clip]{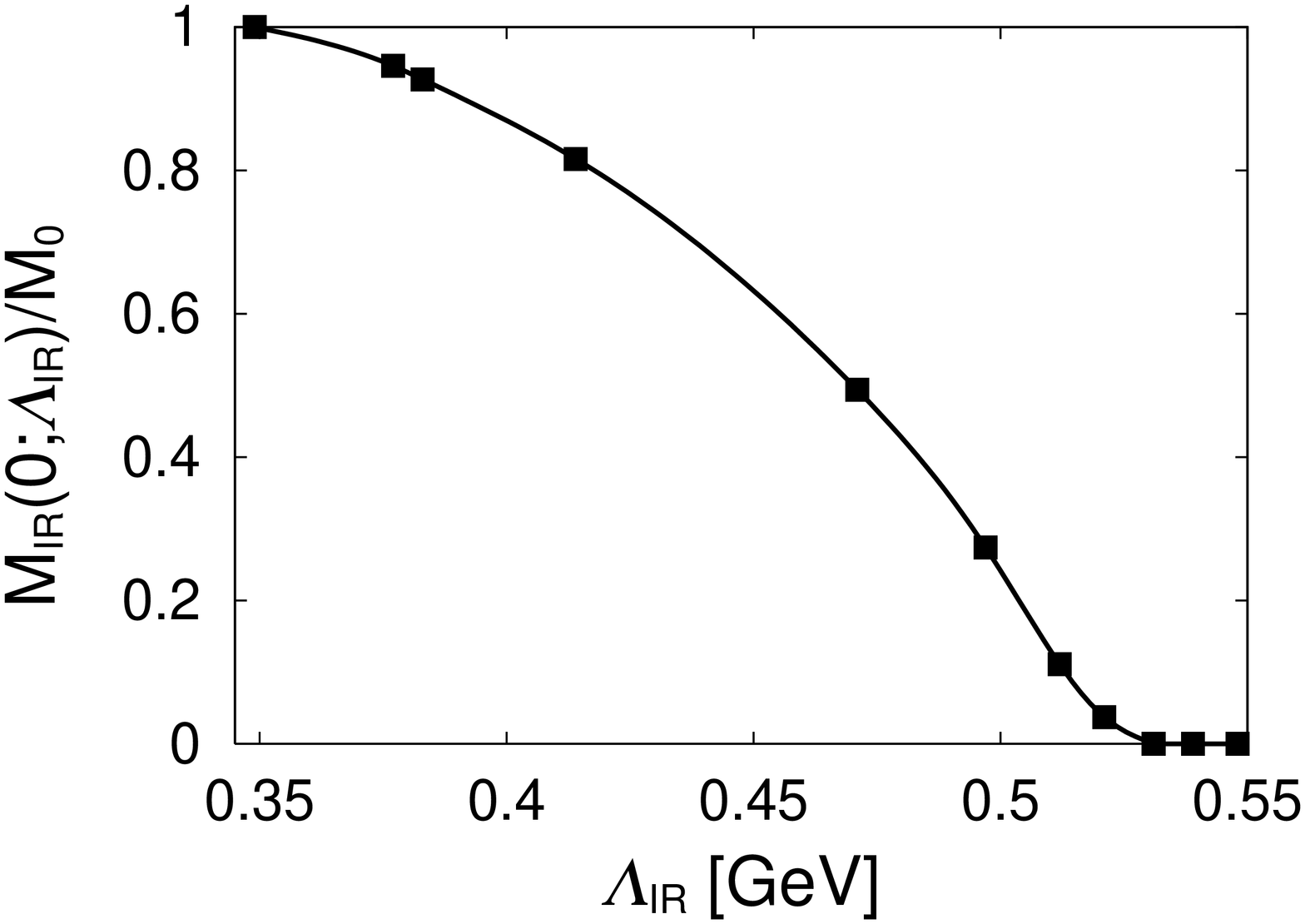}
	\includegraphics[width=6cm,clip]{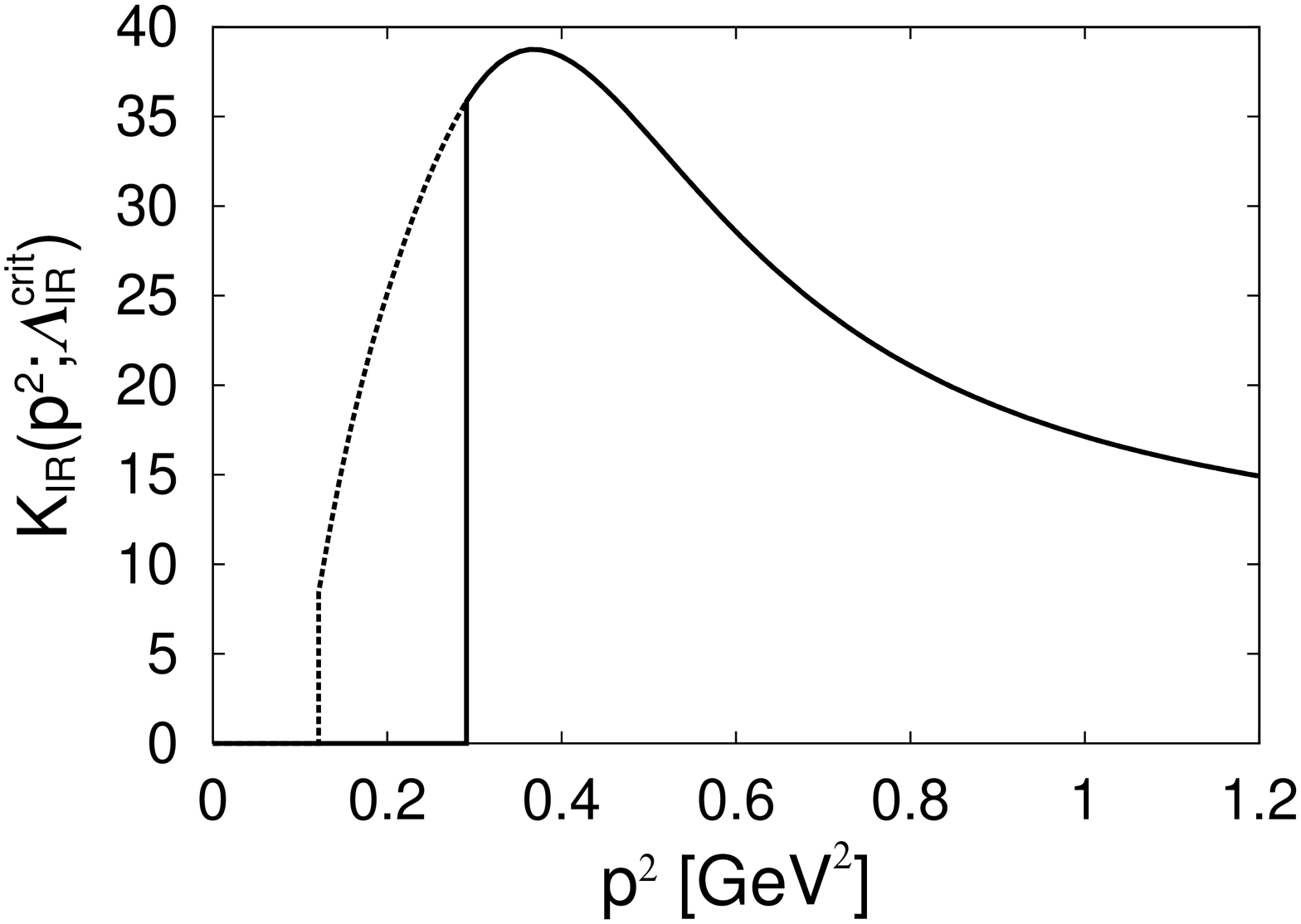}}
\end{center}
\caption{
(a) The infrared quark mass $M_{\rm IR}(0;\Lambda_{\rm IR})/M_0$ in the IR-cut SD equation 
plotted against the artificial IR-cutoff parameter $\Lambda_{\rm IR}$. 
For $\Lambda_{\rm IR}<$0.4GeV, there is no significant effect observed for DCSB.
For $\Lambda_{\rm IR} > \Lambda_{\rm IR}^{\rm crit} \simeq 0.53{\rm GeV}$, 
no DCSB is observed as $M_{\rm IR}(p^2;\Lambda_{\rm IR})=0$.
(b) The IR-cut SD kernel $K_{\rm IR}(p^2;\Lambda_{\rm IR}^{\rm crit})$ for the critical case on DCSB.
The dotted curve denotes the original SD kernel function $K(p^2)$.
}
\label{IRcutcase}
\end{figure}

\begin{figure}
\begin{center}
 \rotatebox{-90}{
	\includegraphics[width=6cm,clip]{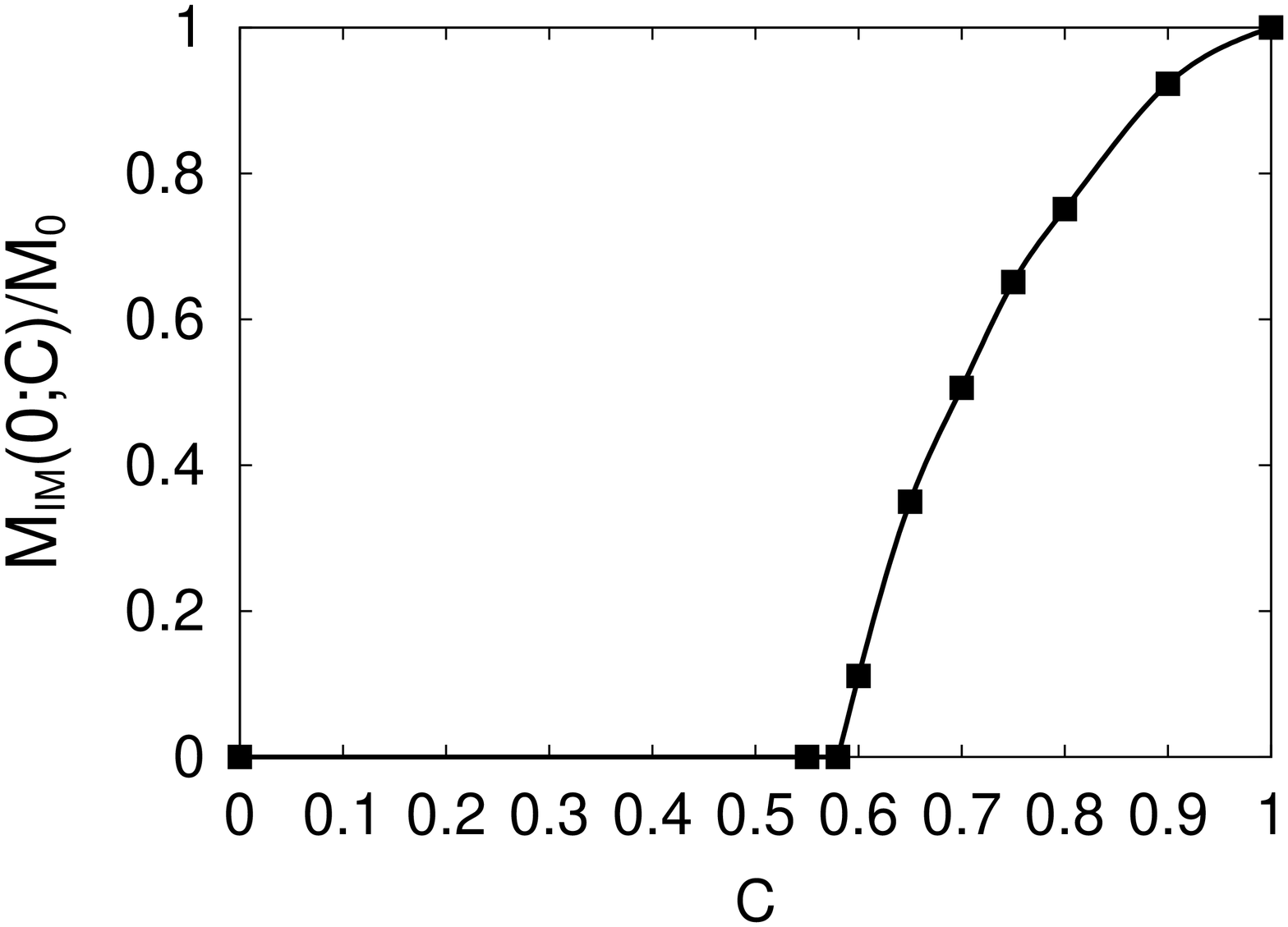}
	\includegraphics[width=6cm,clip]{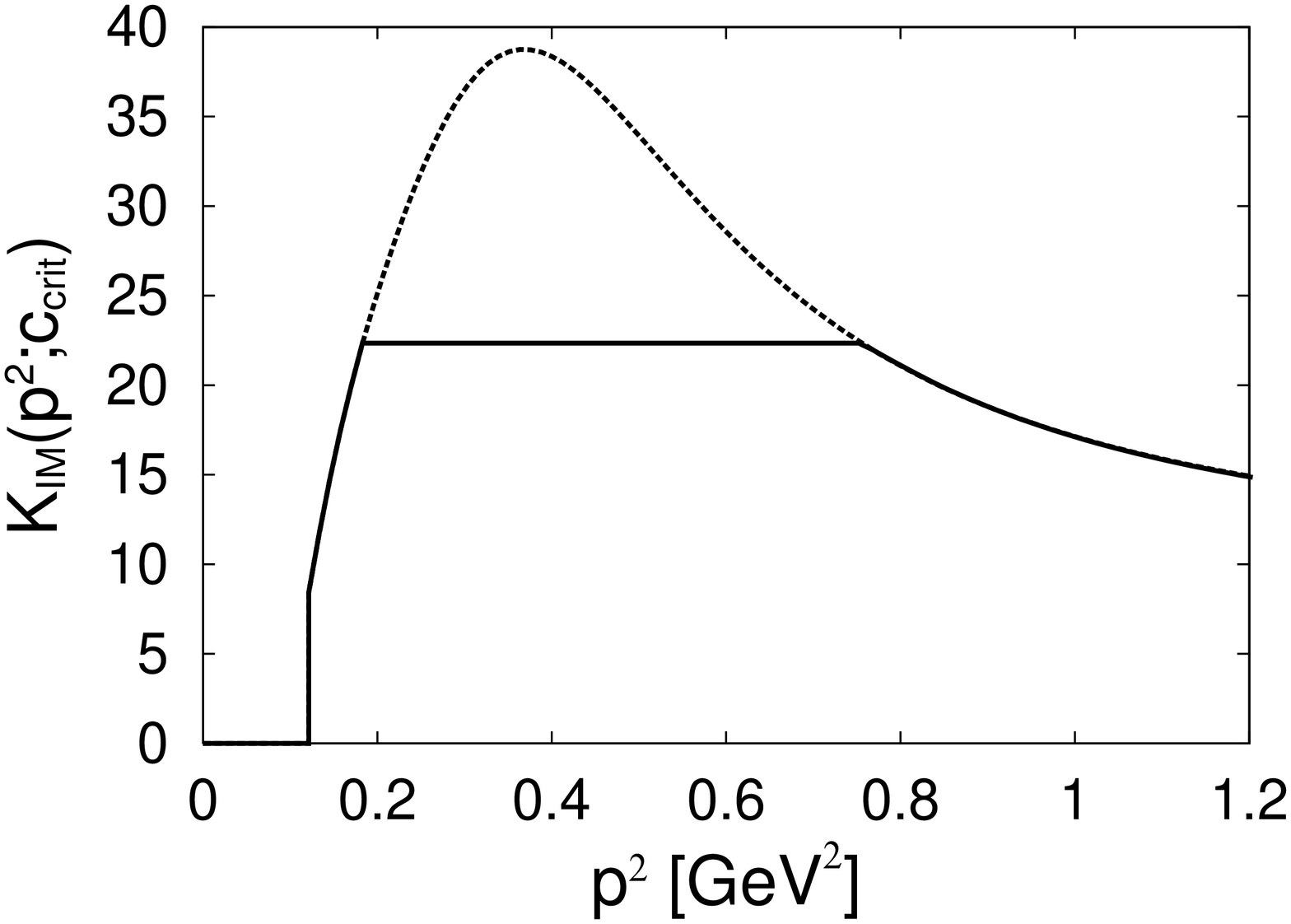}}
\end{center}
\caption{
(a) The infrared quark mass $M_{\rm IM}(0;c)/M_0$ in the IM-suppressed SD equation 
plotted against the artificial IM-suppression parameter $c$.
No DCSB is found as $M_{\rm IM}(p^2;c)=0$ for $c<c_{\rm crit} \simeq 0.58$.
(b) The IM-suppressed SD kernel $K_{\rm IM}(p^2;c_{\rm crit})$ for the critical case on DCSB.
The dotted curve denotes the original SD kernel function $K(p^2)$.
}
\label{IMsuppressedcase}
\end{figure}

\begin{figure}
\begin{center}
 \rotatebox{-90}{
	\includegraphics[width=6cm,clip]{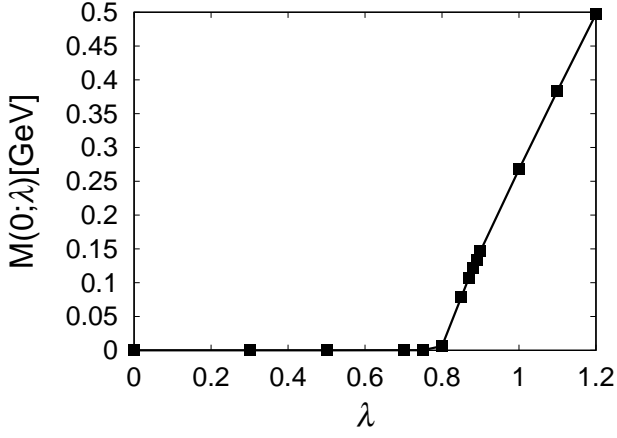}}
\end{center}
\caption{
The infrared quark mass $M(0;\lambda)$ in the scaled SD equation against $\lambda$.
For $\lambda \ge 0.8$, the quark mass function $M(p^2;\lambda)$ is an increasing function on $\lambda$ for fixed value of $p^2$.
For $\lambda < \lambda_{\rm crit} \simeq 0.8$, no DCSB occurs as $M(p^2; \lambda) = 0$.
}
\label{Scaledcase}
\end{figure}

\begin{figure}
\begin{center}
 \rotatebox{-90}{
	\includegraphics[width=6cm,clip]{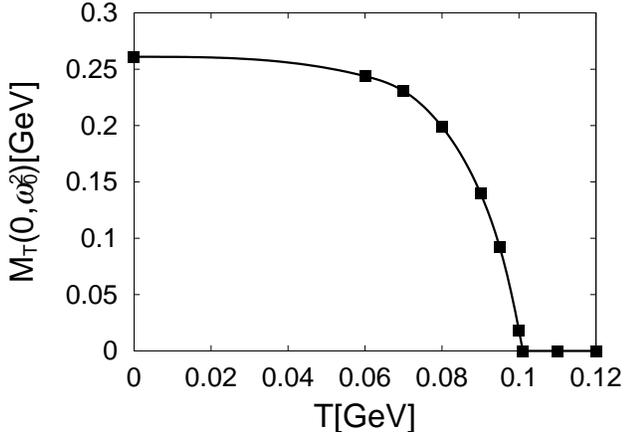}}
\end{center}
\caption{
The thermal infrared quark mass $M_T({\bf p}^2=0,\omega_0^2)$ obtained from the thermal SD equation 
plotted against the temperature $T$.
Chiral symmetry restoration is found at a critical temperature $T_c \simeq 100{\rm MeV}$.
}
\label{ThermalIRQuarkMass}
\end{figure}

\begin{figure}
\begin{center}
 \rotatebox{-90}{
	\includegraphics[width=6cm,clip]{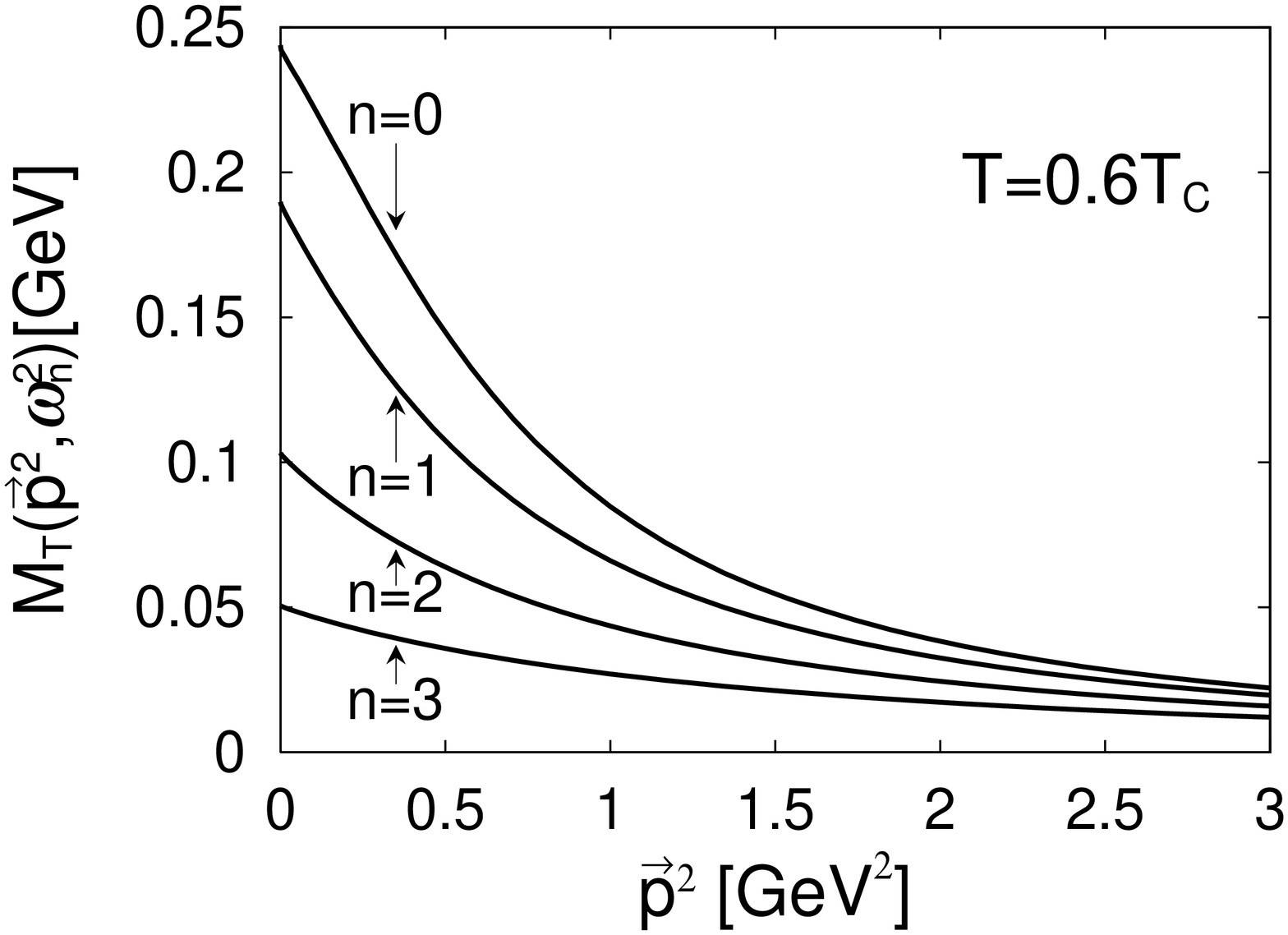}
	\includegraphics[width=6cm,clip]{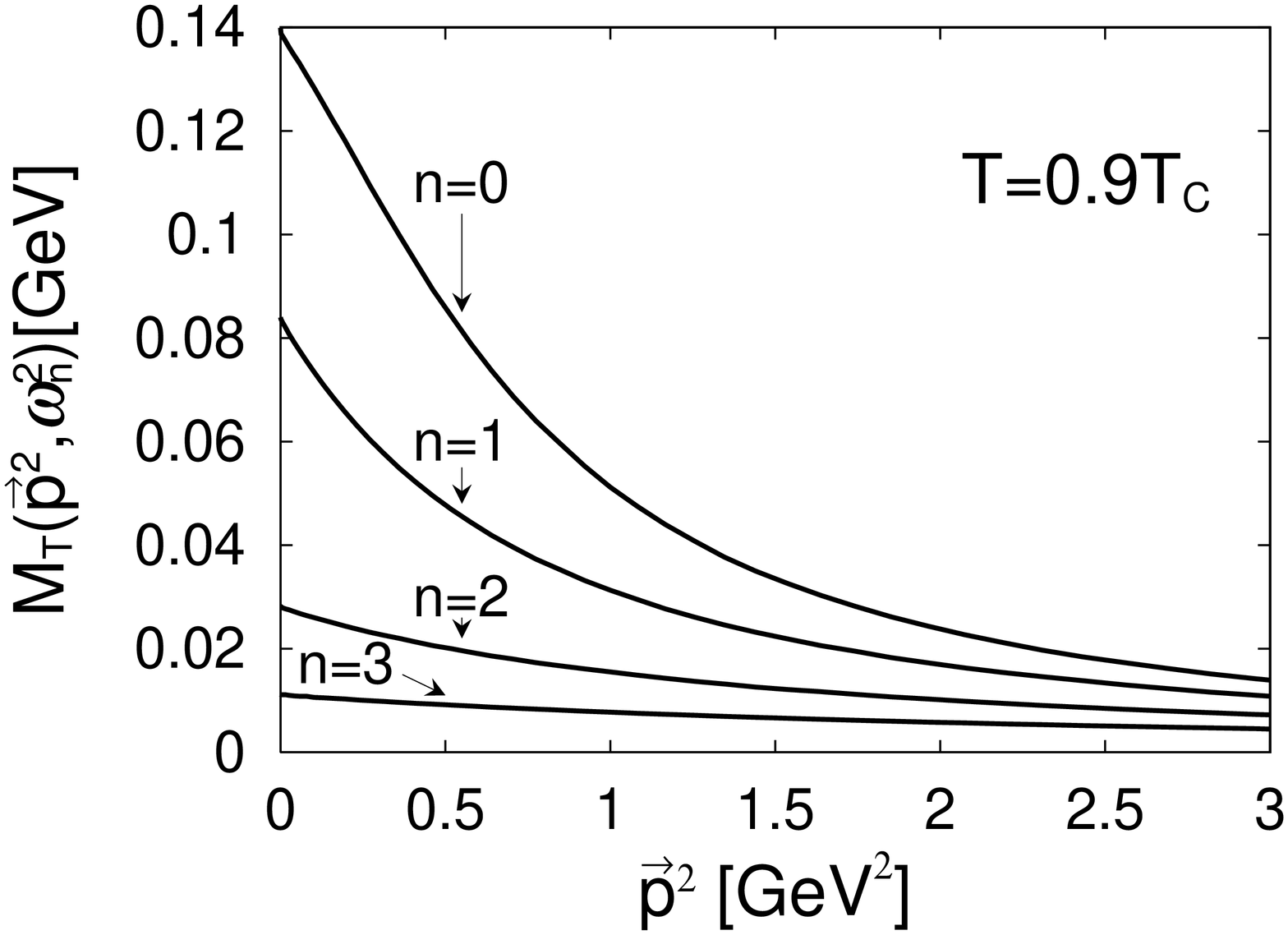}}
\end{center}
\caption{
The thermal quark mass function $M_T({\bf p}^2,\omega_n^2)$ at 
various Matsubara frequencies $n=0,1,2,3$ obtained from the thermal SD equation for 
(a) a low temperature case of $T=0.6 T_c=60{\rm MeV}$ and (b) a high temperature case of $T=0.9T_c = 90{\rm MeV}$.
}
\label{ThermalQuarkMassFunction}
\end{figure}

\begin{figure}
\begin{center}
 \rotatebox{-90}{
	\includegraphics[width=6cm,clip]{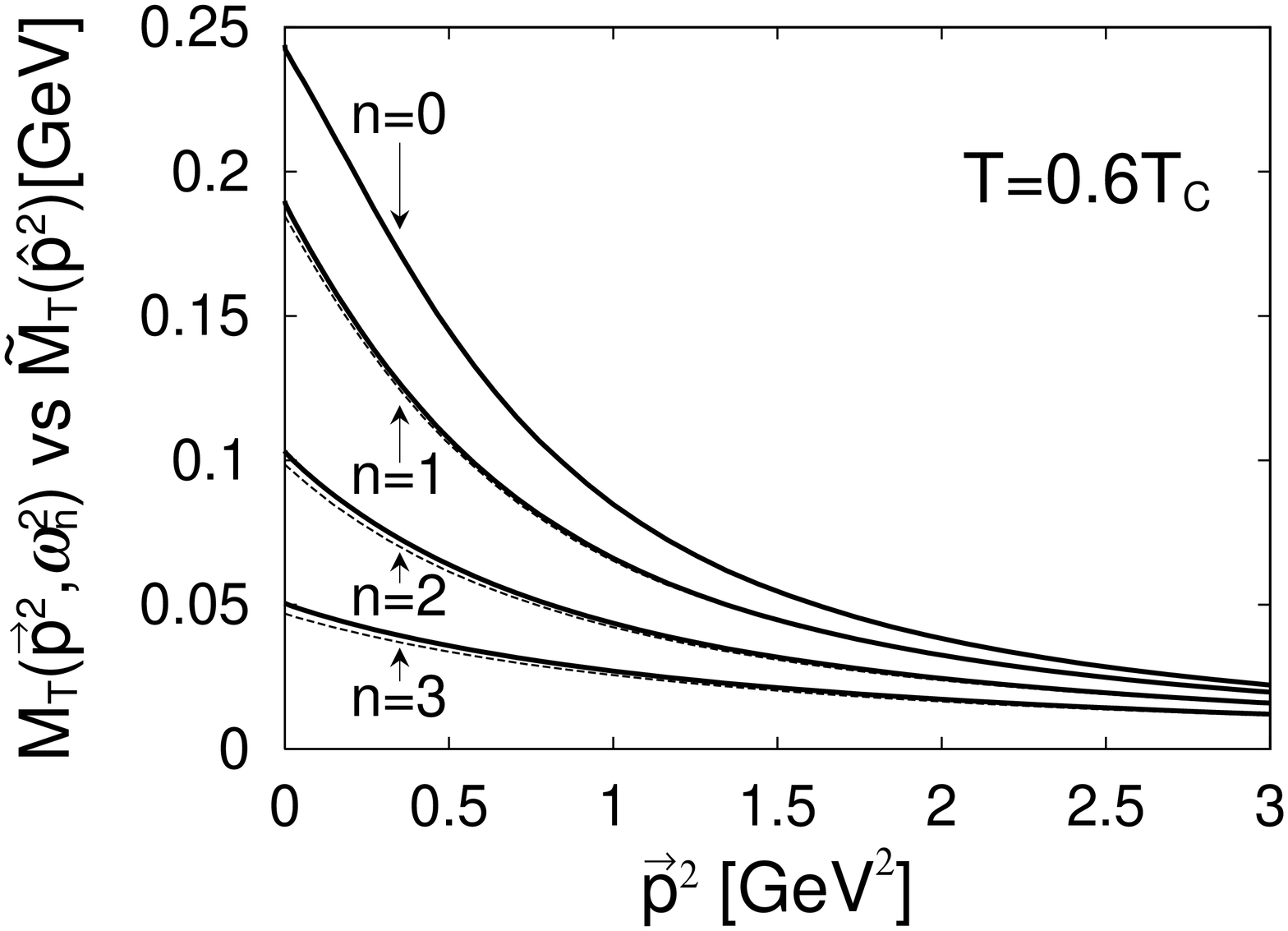}
	\includegraphics[width=6cm,clip]{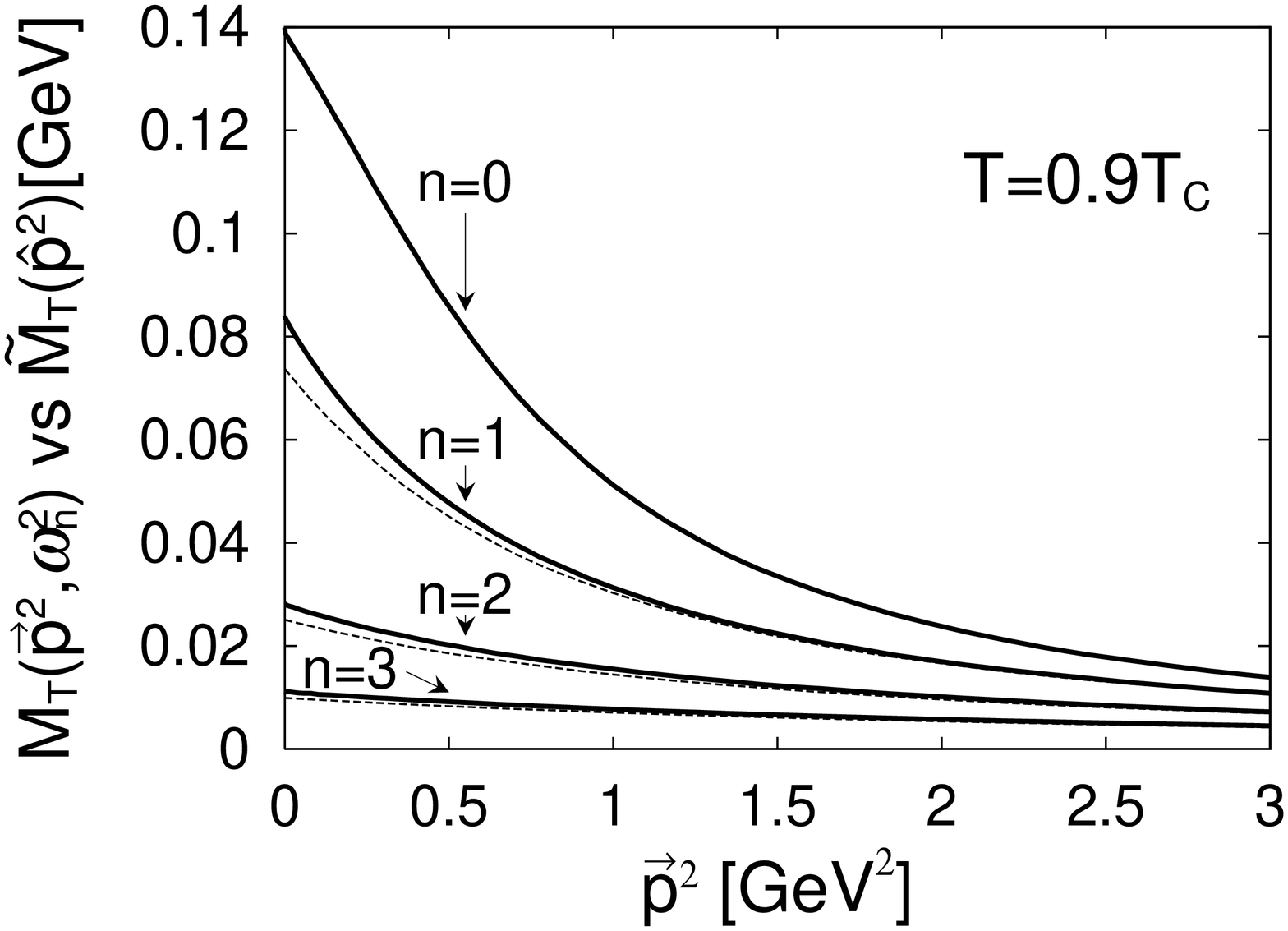}}
\end{center}
\caption{
The comparison of the thermal quark mass function 
$M_T({\bf p}^2,\omega_n^2)$ with $\tilde M_T(\hat p^2)=\tilde M_T({\bf p}^2+\omega_n^2)$ for $n=1,2,3$ at 
(a) a low temperature $T=0.6 T_c=60{\rm MeV}$ and (b) a high temperature $T=0.9T_c = 90{\rm MeV}$.
For each $n$, the dashed curve denotes $\tilde M_T(\hat p^2)$ defined from $M({\bf p}^2,\omega_0^2)$.
}
\label{CovariantLikeRelation}
\end{figure}

\begin{figure}
\begin{center}
 \rotatebox{-90}{
	\includegraphics[width=6cm,clip]{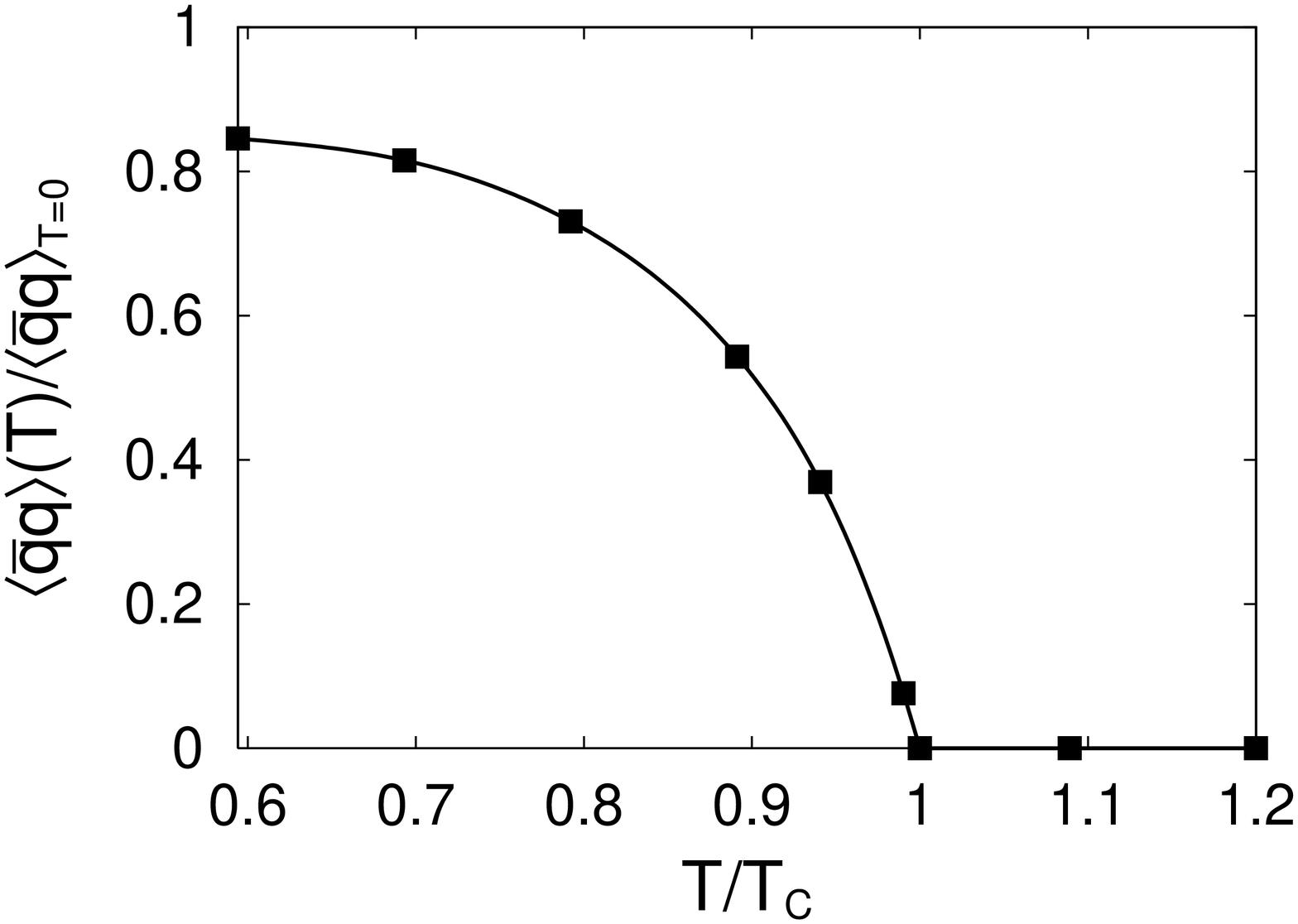}}
\end{center}
\caption{
The thermal quark condensate $\langle \bar qq(T) \rangle_\Lambda/\langle \bar qq(0) \rangle_\Lambda$
plotted against the temperature $T$ for $\Lambda=4{\rm GeV}$.
}
\label{ThermalQuarkCondensate}
\end{figure}


\begin{references}

\bibitem{MGS} For instance, T.~Muta, {\it Foundations of Quantum Chromodynamics} (World Scientific, Singapore, 1998) p.1;
W.~Greiner and A.~Sch\"afer, {\it Quantum Chromodynamics} (Springer-Verlag, Berlin, 1994) p.1.
\bibitem{conf2000} For instance, articles in {\it Quantum Chromodynamics and Color Confinement},  
edited by H.~Suganuma, M.~Fukushima, and H.~Toki (World Scientific, 2001).
\bibitem{NJL61} Y.~Nambu and G.~Jona-Lasinio, Phys.~Rev.~{\bf 122}, 345 (1961).
\bibitem{H8491} K.~Higashijima, Phys. Rev. D{\bf 29}, 1228 (1984); Prog.~Theor.~Phys. Suppl. {\bf 104}, 1 (1991).
\bibitem{M8393} V.~A.~Miransky, Sov.~J~.~Nucl.~Phys.~{\bf 38}(2), 280 (1983); 
{\it Dynamical Symmetry Breaking in Quantum Field Theories}, (World Scientific, Singapore, 1993) p.1.
\bibitem{HT02} For a recent article, M.~Hashimoto and M.~Tanabashi, hep-ph/0210115 (2002) and references threin. 
\bibitem{AKM9091} K.-I.~Aoki, M.~Bando, T.~Kugo, M.~G.~Mitchard, and H.~Nakatani, Prog.~Theor.~Phys. {\bf 84}, 683 (1990);
K-I.~Aoki, T.~Kugo, and M.~G.~Mitchard, Phys.~Lett. {\bf B266}, 467 (1991).
\bibitem{PDG02} Particle Data Group, Phys. Rev. D{\bf 66}, 010001 (2002).
\bibitem{CL84} For instance, T.P.~Cheng and L.F.~Li, 
{\it Gauge Theory of Elementary Particle Physics}, (Clarendon press, Oxford, 1984) p.1.
\bibitem{BKY88} M.~Bando, T.~Kugo, and K.~Yamawaki, Phys.~Rep.~{\bf 164}, 217 (1988) and references therein.
\bibitem{GL82} J.~Gasser and H.~Leutwyler, Phys.~Rep.~{\bf 87}, 77 (1982).
\bibitem{HK94} T.~Hatsuda and T.~Kunihiro., Phys. Rep. {\bf 247}, 221 (1994) and references therein.
\bibitem{DP84} D.~Diakonov and V.~Petrov, Nucl. Phys. {\bf B245}, 259 (1984).
\bibitem{G82} For instance, H.~Georgi, {\it Lie Algebras in Particle Physics} (Benjamin/Cummings, Menlo Park, 1982) p.1.
\bibitem{SST95} H.~Suganuma, S.~Sasaki, and H.~Toki, Nucl. Phys. {\bf B435}, 207 (1995);
S.~Sasaki, H.~Suganuma, and H.~Toki, Prog. Theor. Phys. {\bf 94}, 373 (1995);
H.~Suganuma, S.~Sasaki, H.~Toki, and H.~Ichie, Prog.~Theor.~Phys.~Suppl. {\bf 120}, 57 (1995).
\bibitem{RW94} C.D.~Roberts, and A.G.~Williams, Prog. Part. Nucl. Phys. {\bf 33}, 477 (1994); 
F.T.~Hawes, C.D.~Roberts, and A.G.~Williams, Phys. Rev. D{\bf 49}, 4683 (1994)
\bibitem{R92MM94} H.J.~Rothe, {\it Lattice Gauge Theories} (World Scientific, Singapore, 1992) p.1;
I.~Montvay and G.~M\"unster, {\it Quantum Fields on a Lattice} (Cambridge University Press, Cambridge, England, 1994), p.1.
\bibitem{BBLWZ00} F.D.R.~Bonnet, P.O.~Bowman, D.B.~Leinweber, A.G.~Williams, and J.M.~Zanotti,
Phys. Rev. D{\bf 64}, 034501 (2001);
F.D.R.~Bonnet, P.O.~Bowman, D.B.~Leinweber, and A.G.~Williams, Phys. Rev. D{\bf 62}, 051501 (2000).
\bibitem{BHW02} P.O.~Bowman, U.M.~Heller, and A.G.~Williams, Phys. Rev. D{\bf 66}, 014505 (2002); 
Nucl. Phys. {\bf B}(Proc. Suppl.){\bf 106}, 820 (2002);
Nucl. Phys. (Proc. Suppl.) {\bf 109A}, 163 (2002).
\bibitem{BBLWZ02}F.D.R.~Bonnet, P.O.~Bowman, D.B.~Leinweber, A.G.~Williams, and J.-b.~Zhang, 
Phys. Rev. D{\bf 65}, 114503 (2002).
\bibitem{PS7980} H.~Pagels and S.~Stokar, Phys.~Rev.~D{\bf 20}, 2947 (1979); D{\bf 22}, 2876 (1980).
\bibitem{KM92} T.~Kugo and M.~G.~Mitchard, Phys.~Lett. {\bf B282}, 162 (1992); Phys.~Lett.{\bf B286}, 335 (1992).
\bibitem{IOS03} H.~Iida, M.~Oka, and H.~Suganuma, Proc. of {\it Lattice 2003}, Tsukuba, Japan, Nucl. Phys. {\bf B} (Proc. Suppl.) in press.
\bibitem{BPRT03} M.S.~Bhagwat, M.A.~Pichowsky, C.D.~Roberts, and P.C.~Tandy, Phys. Rev. C{\bf 68}, 015203 (2003). 
\bibitem{SST96} S.~Sasaki, H.~Suganuma, and H.~Toki, Phys. Lett. {\bf B387}, 145 (1996).
\bibitem{ISM02} G.~Boyd, J.~Engels, F.~Karsch, E.~Laermann, C.~Legeland, M.~Lutgemeier, and B.~Petersson, 
Nucl.  Phys. {\bf B469}, 419 (1996); N.~Ishii, H.~Suganuma, and H.~Matsufuru, Phys. Rev. D{\bf 66}, 094506 (2002).
\bibitem{AS99} K.~Amemiya and H.~Suganuma, Phys. Rev. D{\bf 60}, 114509 (1999).

\end{references}
\end{document}